\documentclass[12pt]{article}
\usepackage{newtxtext,newtxmath}
\usepackage[letterpaper,margin=1in]{geometry}
\usepackage{graphicx} 
\usepackage{bm} 
\usepackage{color} 
\usepackage{mathtools}
\usepackage{pdfpages}

\makeatletter
\define@key{Gin}{artifact}[true]{}
\makeatother
\renewenvironment{abstract}
{\quotation}
{\endquotation}
\date{}

\makeatletter
\renewcommand{\fnum@figure}{\textbf{Figure \thefigure}}
\renewcommand{\fnum@table}{\textbf{Table \thetable}}
\makeatother

\usepackage{scicite}

\usepackage{url}

\def\scititle{Dual Topological Channels in a Programmable Mechanical Metamaterial}
\title{\bfseries \boldmath \scititle}

\author{
	Soroush Soltani$^{1}$,
	Jihong Ma$^{1,2,3\ast}$\and
	\small$^{1}$Department of Mechanical Engineering, University of Vermont, Burlington, Vermont, U.S.A.\and\
	\small$^{2}$Department of Physics, University of Vermont, Burlington, Vermont, U.S.A.\and
	\small$^{3}$Materials Science Program, University of Vermont, Burlington, Vermont, U.S.A.\and
	\small$^\ast$Corresponding author. Email: Jihong.Ma@uvm.edu
}

\begin{document} 
	\maketitle
\begin{abstract} \bfseries \boldmath
\noindent Topological mechanical metamaterials generally derive their robustness from one of two distinct mechanisms: kinematic topology, which localizes zero-frequency floppy modes through geometric compatibility, or band topology, which localizes finite-frequency waves through topological bandgaps. Because these mechanisms arise from fundamentally different physical principles, mechanical systems are typically engineered to exploit either static or dynamic topological functionality, but not both simultaneously. Here we demonstrate that a single mechanical metamaterial can host two independent topological channels governing static and dynamic response within the same architecture. Using a generalized rotor-chain lattice with two coupled rotational degrees of freedom per unit cell, we realize the coexistence of a topological polarization mode and a finite-frequency topological band mode. We show that the two channels are governed by distinct invariants and can be programmed independently through geometry, angular asymmetry, and stiffness dimerization. Remarkably, despite being governed by distinct topological invariants and controlling different physical responses, the two channels undergo topological transitions at a common symmetry-controlled critical geometry. Experimental measurements using scanning laser Doppler vibrometry confirm simultaneous localization of a boundary floppy mode and a finite-frequency domain-wall state in the same structure. These results establish dual topological channels as a general design principle for multifunctional mechanical metamaterials and demonstrate how static deformation and dynamic wave transport can be programmed independently within a single architecture.

\end{abstract}	
\noindent	
\section{Introduction} 
    Mechanical metamaterials that exploit topological principles, first developed in the context of electronic topological insulators~\cite{Hasan1}, can localize, guide, and protect mechanical energy through bulk invariants that are insensitive to many forms of disorder ~\cite{tang1}. These capabilities have established topology as a powerful framework for designing robust mechanical response, including vibration isolation, waveguiding, and programmable motion~\cite{Bertoldi1,Ni1,Cummer1,ma1,Ma2}. Beyond achieving robustness, an emerging challenge for the field is the realization of multifunctional architectures in which multiple mechanical functions can coexist and be independently controlled within a single material platform. Such capabilities could enable materials that simultaneously regulate static deformation, dynamic wave transport, signal routing, and adaptive mechanical response.
    
    Topological mechanical metamaterials have largely developed along two distinct directions. The first exploits \textbf{kinematic topology}, which arises in isostatic or Maxwell lattices whose constraint geometry supports zero-frequency floppy modes governed by a topological polarization determined entirely by lattice compatibility~\cite{Kane1,Lubenski1,Mao1,Maxwell1,Jacobs1,Calladine1}. In these systems, topology is encoded in geometric constraints and exists even in the absence of elastic energy. This mechanism has been realized in a variety of architectures, including origami-inspired structures and reconfigurable kagome lattices, where protected boundary modes emerge from geometry alone~\cite{Chen2,Rocklin1}.
    
    A second direction exploits \textbf{dynamic topology}, encoded in the topology of finite-frequency band structures. Mechanical analogs of the Su--Schrieffer--Heeger (SSH) model and related phononic topological systems employ alternating couplings or geometrical asymmetries to create bandgaps that host protected domain-wall and edge states~\cite{Ma2,su1,su2,Pal1,Rajabpoor1,Albaba1}. Unlike kinematic topology, these phenomena are fundamentally governed by elastic interactions and wave propagation, and therefore depend on both geometry and constitutive coupling strengths. Dynamical topology has enabled robust manipulation of elastic-wave propagation and established topological band engineering as a powerful strategy for controlling mechanical energy transport.

    Although both paradigms produce robust boundary localization, they arise from fundamentally different descriptions of mechanics. Kinematic topology is encoded in compatibility relations and geometric constraints, whereas dynamical topology emerges from stiffness-dependent wave physics. As a result, these concepts have largely evolved as separate design strategies. Structures hosting topological floppy modes are typically engineered to manipulate quasi-static mechanical response, while systems hosting finite-frequency topological states are generally optimized for wave transport and energy localization. Consequently, topological metamaterials are often designed to exploit either static or dynamic functionality, but rarely both simultaneously. More broadly, the field lacks a framework for realizing multiple independently controllable topological functionalities within a single mechanical architecture.

    This separation is particularly evident in rotor-chain systems. The classical rotor chain hosts a zero-frequency topological polarization mode governed by Maxwell compatibility and serves as an iconic realization of kinematic topology~\cite{Mao1}. However, because the unit cell contains only a single rotational degree of freedom, it provides no independent channel through which finite-frequency band topology can be introduced without simultaneously altering the geometry responsible for the polarization invariant. More recently, several studies have shown that introducing elastic restoring forces into Maxwell lattices can lift topological floppy modes to finite frequency while preserving their underlying topological character~\cite{Ma3,sun1,zhang1}. Related finite-frequency Maxwell modes have also been realized in self-dual kagome geometries~\cite{danawe1}. These developments reveal that kinematic topology can survive beyond the strict zero-frequency limit, but they still describe the persistence of a single topological invariant across different frequency regimes rather than the coexistence of multiple independently switchable topological channels within the same structure. 
    
    A further challenge is conceptual. Conventional descriptors of finite-frequency topology, including the Zak phase and related symmetry-based formulations, are most naturally defined in systems possessing inversion or chiral symmetry~\cite{Zak1,Li1,jiao1,Wang1}, whereas creating a finite-frequency channel in a non-centrosymmetric lattice often requires introducing internal asymmetries and extra degrees of freedom that break these symmetries. Whether distinct kinematic and dynamical topology can therefore coexist as distinct invariants within the same non-centrosymmetric mechanical architecture, while remaining independently programmable, remains an open question. Likewise, it is unknown whether the corresponding topological transitions arise from unrelated physical mechanisms or whether they can be connected through a common symmetry principle.

    Here, we address these questions using a generalized rotor-chain metamaterial that provides the minimal realization of \textbf{dual topological channels}. By introducing a second rotational degree of freedom into the classical rotor chain, we create a lattice that simultaneously supports a topological polarization mode governing static response and a finite-frequency topological state governing dynamic response. Using analytical theory, numerical calculations, and experimental measurements, we show that the two channels are characterized by distinct topological invariants and remain independently programmable: the kinematic channel is governed solely by the average pendulum orientation, whereas the dynamical channel can be programmed independently through equilibrium-angle asymmetry, axial-stiffness dimerization, or rotational-stiffness dimerization. To characterize the finite-frequency channel in the absence of exact chiral symmetry, we introduce a projected winding-number formulation that remains well-defined in the non-centrosymmetric regime~\cite{Mochizuki1,Wang2}. Remarkably, despite their different physical origins, we find that the two topological channels undergo transitions at a common inversion-symmetry-controlled critical geometry. Experimental measurements using scanning laser Doppler vibrometry confirm the simultaneous existence of a boundary-localized floppy mode and a finite-frequency domain-boundary state within the same architecture. Together, these results establish dual topological channels as a design principle for multifunctional mechanical metamaterials and demonstrate how static deformation and dynamic wave transport can be independently programmed within a single material system.
	
	\section{Results}
\subsection{A minimal platform for dual topological channels}

\begin{figure}[h] 
	\centering
	\includegraphics[width=1\textwidth]{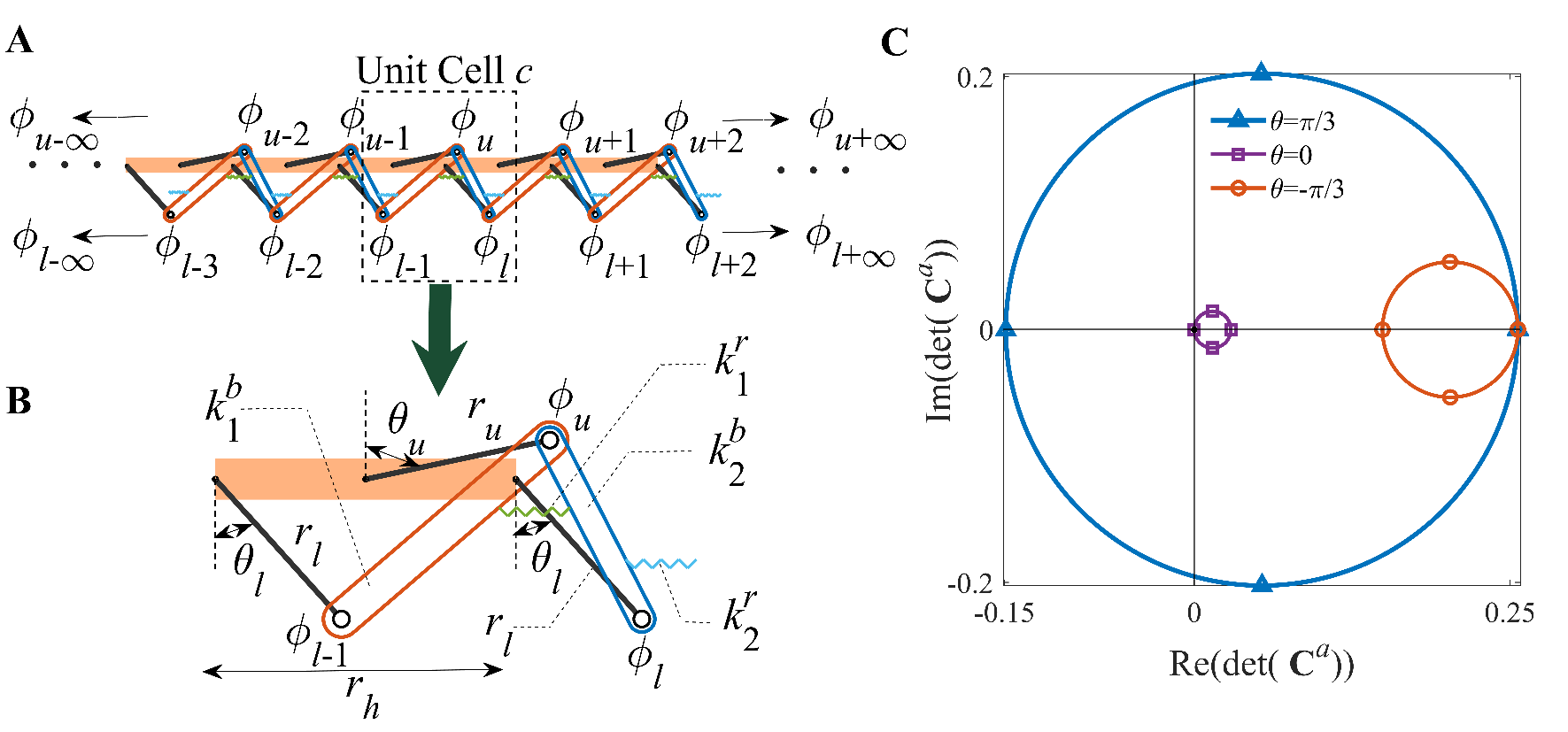} 
	\caption{\textbf{Unit cell and zero-frequency invariant.} (\textbf{A}) Infinite zigzag rotor chain built from repeating unit cells $c$, each with independent upper ($\phi_u$) and lower ($\phi_l$) pendulum rotations. (\textbf{B}) Geometric and stiffness parameters of a single unit cell. (\textbf{C}) Winding path of $\det(\mathbf{C}^a)$ at $\theta_u=\theta_l=\theta=-\pi/3,0,\pi/3$, recovering the classical single-degree-of-freedom rotor chain in this symmetric limit.} 
	\label{Fig01} 
\end{figure}

Each unit cell of our metamaterial consists of two rigid pendulums --- of length $r_u$ (upper) and $r_l$ (lower) --- pivoting about out-of-plane hinges and linked by a rigid horizontal ribbon of width $r_h$ (Fig.~\ref{Fig01}, A and B). In equilibrium, the pendulums sit at angles $\theta_u$ and $\theta_l$ from vertical. Zigzag bars with axial stiffnesses $k_1^b$ and $k_2^b$, together with rotational hinge springs $k_1^r$ and $k_2^r$, couple each pendulum's rotation to its neighbors. Unlike the classical rotor chain, which possesses only a single rotational degree of freedom per unit cell~\cite{Mao1}, the present architecture has two independent rotational coordinates, $\phi_u$ and $\phi_l$. This additional internal degree of freedom fundamentally expands the topological design space of the rotor chain by creating distinct pathways through which geometry and elasticity can be programmed.

Under Bloch–Floquet boundary conditions, the extensions of the two bars and the relative rotations at the two hinges are related to the pendulum rotations through a compatibility matrix $\widetilde{\mathbf{C}}^p(qa)$, and the corresponding Bloch stiffness matrix $\widetilde{\mathbf{K}}^p(qa) = (\widetilde{\mathbf{C}}^p)^T \widetilde{\mathbf{D}} \widetilde{\mathbf{C}}^p$, with $\widetilde{\mathbf{D}} = \text{diag}(k^b_1, k^b_2, k^r_1, k^r_2)$, governs bulk wave propagation (complete derivation in Supplementary Text S1).  

Importantly, this formulation reveals two distinct topological channels embedded within the same lattice. The first governs static response through a zero-frequency floppy mode localized by a kinematic invariant $n$, while the second governs dynamic response through finite-frequency wave localization protected by a dynamical invariant $w$. Because the two channels arise from different physical principles, they can in principle be manipulated independently while coexisting within a single mechanical architecture.

The zero-frequency polarization $n$ is determined by the winding of $\det(\mathbf{C}^a)$, the axial sub-block of the compatibility matrix (as detailed in Supplementary Text S1 and S6), around the origin as the Bloch wavevector traverses the Brillouin zone. The rotational-hinge extensions, by contrast, trace a path that stays tangent to the origin and thus do not contribute to this winding, leaving $n$ set purely by the axial geometry. Formally,
\begin{equation}\label{eq:winding_zero}
	n = \frac{1}{2\pi \textrm{i}} \int_0^{2\pi/a} \text{d}(qa) \frac{\text{d}}{\text{d}(qa)} \ln \det [\mathbf{C}^a(qa)],
\end{equation}
where $n=1$ and $n = 0$ correspond, respectively, to right- and left-edge localization of the associated zero-frequency floppy mode.

The invariant $n$ therefore characterizes a purely kinematic topological channel that depends only on lattice geometry and remains independent of the stiffness distribution. In the symmetric limit $\theta_u=\theta_l=\theta$, $k^r_1=k^r_2=0$, and $k^b_1=k^b_2$, the model reduces to the classical single-degree-of-freedom rotor chain~\cite{Mao1}: the trajectory of $\det(\mathbf{C}^a)$ encircles the origin ($n = 1$) for $\theta =\pi/3$ and does not ($n = 0$) for $\theta=-\pi/3$ (Fig.~\ref{Fig01}C), recovering the known zero-frequency topological transition (Fig.~\ref{Fig01}C).

The same lattice also supports a second, fundamentally different topological channel. Whereas the first is encoded in compatibility relations and governs static response, the second emerges from the topology of finite-frequency wave propagation and governs the existence of localized vibrational states within a bulk bandgap. We decompose the $2 \times 2$ Bloch stiffness sub-block $\widetilde{\mathbf{K}}^p(qa)$ in the Pauli basis $\boldsymbol{\sigma} = (\sigma_x, \sigma_y, \sigma_z)$:
\begin{equation}\label{eq:pauli}
	\widetilde{\mathbf{K}}^p(qa) = d_0\sigma_0 + d_x(qa)\sigma_x + d_y(qa)\sigma_y + d_z\sigma_z.
\end{equation}
In standard SSH models with exact chiral symmetry, the mass term $d_z$ vanishes; in our non-centrosymmetric model, however, it takes the finite value $d_z = k^b(\delta_0 - \delta_1)/2$. Because this term arises purely from the local, asymmetric geometry of the unit cell and is independent of the momentum $qa$, it represents a static energy offset that does not couple to wave propagation and can therefore be projected out along with the average spectral shift $d_0\sigma_0$ (Supplementary Text S2–S5). Subtracting both terms leaves the dynamically isolated effective chiral Hamiltonian $\widetilde{\mathbf{K}}'_{\rm eff}(qa) = d_x(qa)\sigma_x + d_y(qa)\sigma_y$, from which the finite-frequency invariant $w$ is evaluated as the integral of the Berry connection:
\begin{equation}\label{eq:winding_finite}
	w = \frac{1}{4\pi \textrm{i}} \int_{-\pi/a}^{\pi/a} \text{tr}\left[\sigma_z \left(\widetilde{\mathbf{K}}'_\text{eff}(qa)\right)^{-1} \partial_q \widetilde{\mathbf{K}}'_\text{eff}(qa)\right]\text{d}q.
\end{equation}
The invariant $w$ therefore remains a robust topological diagnostic despite the loss of exact chiral symmetry --- a conclusion we corroborate numerically by tracking the accumulated phase difference between the two pendulum eigenvector components across the Brillouin zone (Supplementary Text S8). Together, $n$ and $w$ establish the presence of dual topological channels within the same unit cell, laying the foundation for independently programmable static and dynamic response.

\subsection{Breaking inversion symmetry  activates a finite-frequency channel while preserving the kinematic channel}

\begin{figure}[h]
	\centering
	\includegraphics[width=0.89\textwidth]{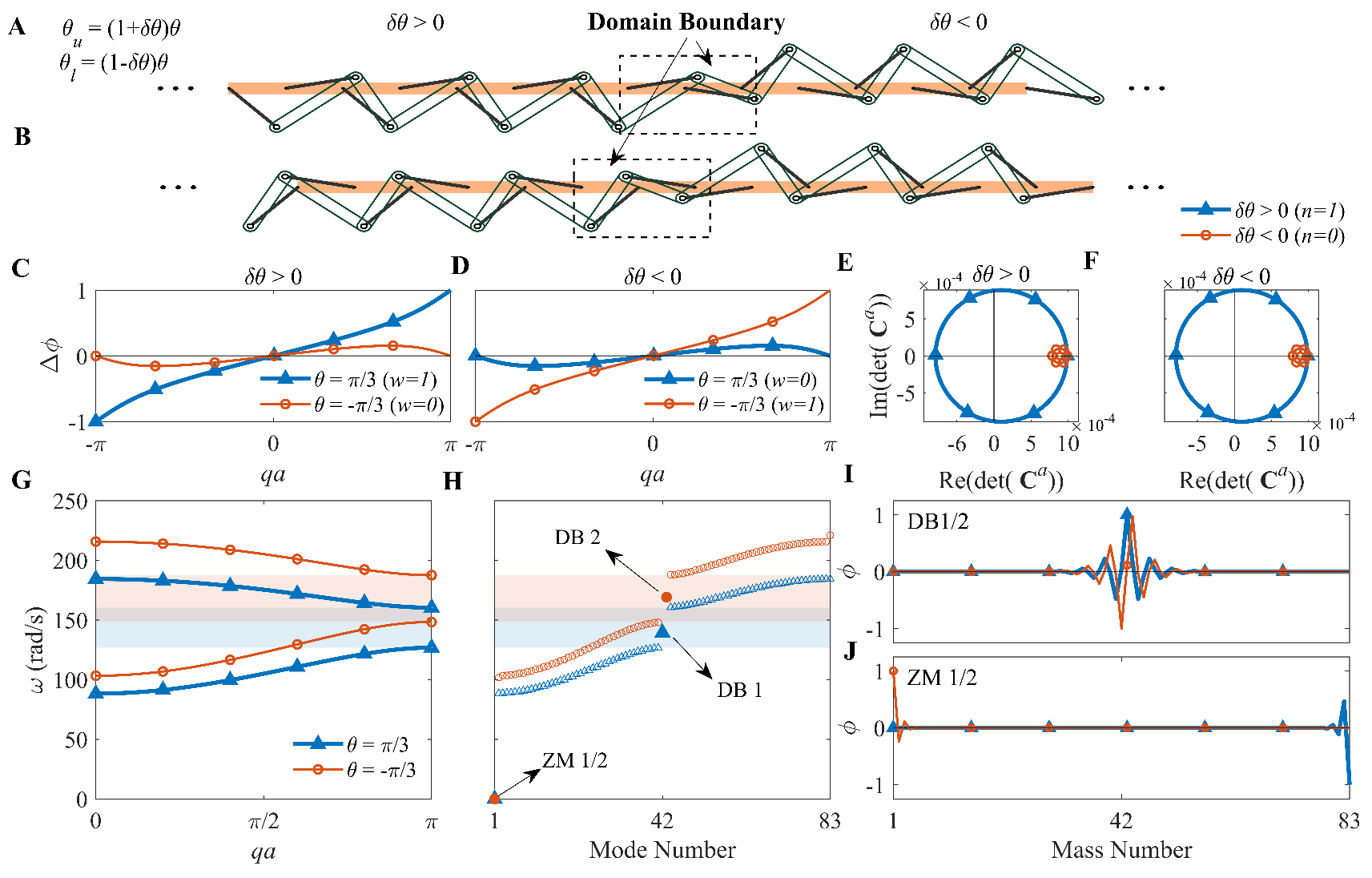}
	\caption{\textbf{Angular asymmetry opens a finite-frequency channel without disturbing zero-frequency topology.} (\textbf{A} and \textbf{B}) Finite lattices with domain boundaries between regions of opposite $\delta\theta$ for $\theta>0$ (\textbf{A}) and $\theta<0$ (\textbf{B}). (\textbf{C} and \textbf{D}) Eigenvector phase difference $\Delta\phi$ confirming the finite-frequency winding number $w$ for $\delta\theta>0$ (\textbf{C}) and $\delta\theta<0$ (\textbf{D}). (\textbf{E} and \textbf{F}) Corresponding polarization winding trajectories of $\det(\mathbf{C}^a)$. (\textbf{G}) Dispersion showing the induced finite-frequency bandgap. (\textbf{H}) Supercell spectrum showing localized domain-boundary (DB) and zero-frequency (ZM) states. (\textbf{I} and \textbf{J}) DB mode shapes for $\theta>0$ (\textbf{I}) and $\theta<0$ (\textbf{J}). (\textbf{K}) ZM mode shapes for the configurations in (\textbf{A}) and (\textbf{B}), showing unchanged ZM topology despite the new finite-frequency gap.}
	\label{Fig02}
\end{figure}

Having identified the two topological channels, we next ask whether they can coexist within a single parameter regime. To do so, we introduce an internal asymmetry between the two pendulums while preserving their average orientation, parameterizing the equilibrium angles as $\theta_u = (1+\delta\theta)\theta$ and $\theta_l = (1-\delta\theta)\theta$, which breaks space-inversion symmetry for any $\delta \theta\neq0$ (here, we set $\delta\theta = \pm0.3$) while leaving the average angle $\theta_{avg}=(\theta_u+\theta_l)/2=\theta$ unchanged. All other parameters, including $k_{1,2}^b$ and $k_{1,2}^r$, remain identical. Two representative scenarios are shown in Fig.~\ref{Fig02}, A ($\theta=\pi/3$) and B ($\theta=-\pi/3$). This asymmetry induced by $\delta \theta\neq0$ opens a finite-frequency bandgap separating the acoustic and optical branches (Fig.~\ref{Fig02}G), whose width increases with $|\delta\theta|$ (Supplementary Fig.~S2 and Fig.~S3). More importantly, it activates a finite-frequency topological channel without altering the geometric quantity that controls the zero-frequency polarization.

The inversion-symmetry breaking introduced by $\delta\theta$ does more than open a bandgap: it also determines the topology of the resulting band structure. Both the projected winding number calculated from Eq.~\ref{eq:winding_finite} and the accumulated eigenvector phase difference $\Delta\phi$ (Fig.~\ref{Fig02}, C and D) show that the finite-frequency channel enters a topological phase whenever $\theta$ and $\delta\theta$ share the same sign and a trivial phase when their signs are opposite. Consequently, reversing the sign of $\delta\theta$ across a domain boundary changes the finite-frequency invariant by one and generates a localized domain-boundary state regardless of the sign of $\theta$ (Fig.~\ref{Fig02}, I). The dynamical equations governing these domain-boundary-containing finite lattices are derived in detail in Supplementary Text S1.

Despite this broken inversion symmetry, the zero-frequency polarization remains unchanged. The winding trajectories of $\det(\mathbf{C}^a)$ reveal that $n$ depends only on the sign of $\theta$ and is completely insensitive to $\delta\theta$ (Fig.~\ref{Fig02}, E and F). Consequently, the floppy mode remains localized at the same boundary selected by the average geometry of the chain (Fig.~\ref{Fig02}J).

The two channels therefore depend on fundamentally different aspects of the same structure. The kinematic channel is governed solely by the average lattice geometry, whereas the dynamical channel emerges from how that geometry is partitioned between the two pendulums. A single unit cell thus hosts two independently controllable topological degrees of freedom, one governing static response and the other governing dynamic response.

\subsection{A shared critical geometry links the two topological channels}
The coexistence demonstrated above raises a deeper question. If the kinematic and dynamical channels are governed by distinct topological invariants and can be programmed independently, are their topological transitions fundamentally unrelated, or do they share a common origin?

This result is not obvious a priori. The kinematic invariant $n$ is determined purely by compatibility and geometric polarization, whereas the dynamical invariant $w$ is derived from finite-frequency wave physics. One therefore might expect the two transitions to occur at unrelated points in parameter space. Instead, analytical evaluation of Eq.~\ref{eq:winding_finite} reveals that the finite-frequency invariant $w$ is ultimately controlled by the relative strength of two geometric coupling pathways connecting the upper and lower pendulums. As shown in Supplementary Text S2, the winding number reduces to a comparison between the products, $\delta_1^{(1)}\delta_1^{(2)}$ and $\delta_2^{(1)}\delta_2^{(2)}$, 
\begin{equation}\label{eq:w_criterion}
	w =
	\begin{cases}
		1 & \text{if } |\delta_1^{(1)}\delta_1^{(2)}| > |\delta_2^{(1)}\delta_2^{(2)}|,\\
		0 & \text{if } |\delta_1^{(1)}\delta_1^{(2)}| < |\delta_2^{(1)}\delta_2^{(2)}|.
	\end{cases}
\end{equation}
Each product represents the effective coupling transmitted through one of the two zigzag bars, allowing the finite-frequency topology to be interpreted as a competition between two geometry-mediated interaction pathways.

The resulting quantity plays a role analogous to the intracell-to-intercell hopping ratio of the SSH model. Importantly, however, no physical stiffness dimerization is introduced here. Both bars possess identical axial stiffness, and the apparent ``dimerization'' emerges entirely from geometry. By assigning different equilibrium angles to the two pendulums ($\theta_u\neq\theta_l$), the lattice redistributes how strongly each bar couples the rotational degrees of freedom, generating an effective topological asymmetry without modifying the material properties of the structure.

We therefore define the \textit{phase-transition ratio} $|\delta_1^{(1)}\delta_1^{(2)}/\delta_2^{(1)}\delta_2^{(2)}|$, which serves as the finite-frequency analogue of the SSH hopping ratio. The transition $w=0\xleftrightarrow{}1$ occurs precisely when this ratio crosses unity.

\begin{figure}[h]
	\hspace*{-15mm}
	\includegraphics[width=1.15\textwidth]{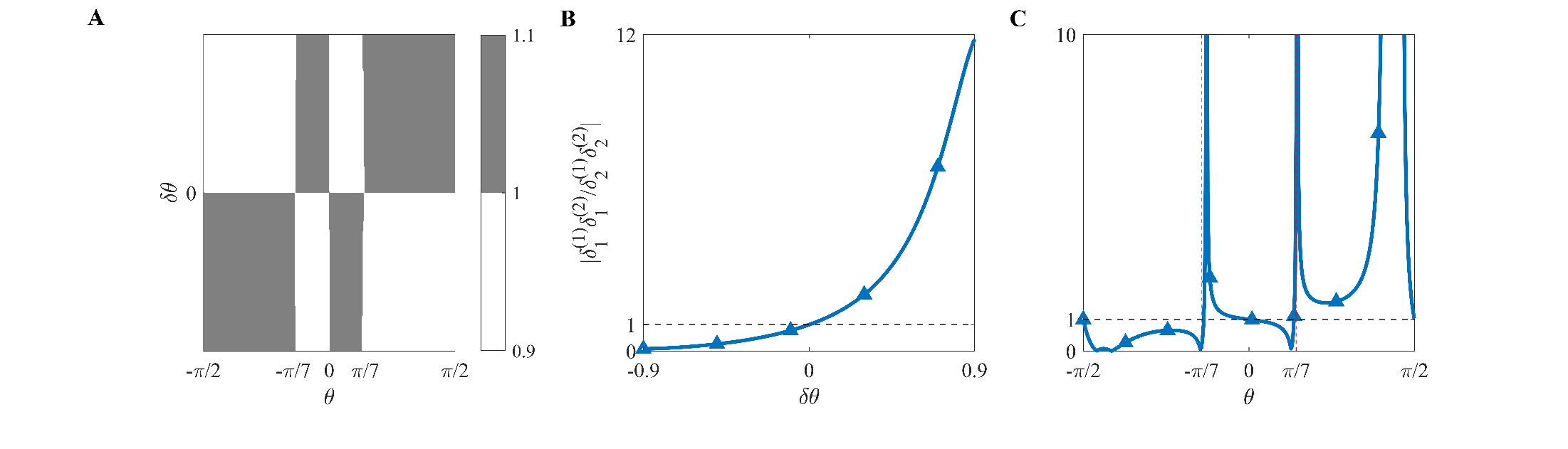}
	\caption{\textbf{Shared critical geometry and the SSH-like phase-transition ratio.} (\textbf{A}) Map of the phase-transition ratio $|\delta_1^{(1)}\delta_1^{(2)}/\delta_2^{(1)}\delta_2^{(2)}|$ relative to unity (gray: $w=1$; white: $w=0$) over $\theta$ and $\delta\theta$; the $\theta=0$ boundary coincides with the zero-frequency transition at $\theta_{avg}=0$ for every $\delta\theta$. (\textbf{B}) The ratio versus $\delta\theta$ at $\theta=\pi/3$. (\textbf{C}) The ratio versus $\theta$ at $\delta\theta=0.3$; the dashed line marks the ratio-equals-1 threshold separating the two finite-frequency phases.}
	\label{Fig03}
\end{figure}

Fig.~\ref{Fig03}A maps this ratio across the $(\theta, \delta\theta)$ parameter space explored in Fig.~\ref{Fig02}; the gray/white boundary marks where the ratio crosses 1, i.e., where $w$ changes character. The most striking feature is that one of the finite-frequency phase boundaries remains locked to $\theta=0$ for all values of $\delta\theta$. This is precisely the same geometry at which the zero-frequency polarization invariant changes sign. At this point the average pendulum orientation vanishes, inversion symmetry is restored, and both the polarization gap and the finite-frequency bandgap close simultaneously. The two channels therefore remain distinct, yet their topological transitions become synchronized through the same underlying symmetry condition.

This result reveals an unexpected relationship between geometric and dynamical topology. The kinematic and dynamical channels originate from different physical principles and are characterized by different invariants, yet both detect the same critical restoration of inversion symmetry. The rotor chain therefore provides a concrete example in which two independently programmable topological channels remain connected through a common symmetry-controlled transition.

This boundary sits at fixed values of $\theta$ ($\theta = 0, \pm\pi/7$), essentially independent of $\delta\theta$ (Fig.~\ref{Fig03}, A and C). Because $\theta_{avg}=\theta$ under the parameterization $\theta_u=(1+\delta\theta)\theta$, $\theta_l=(1-\delta\theta)\theta$ used here, the $\theta=0$ boundary is exactly the point at which the polarization invariant $n$ also transitions: the unit cell regains inversion symmetry, and the zero-frequency polarization gap and the finite-frequency bulk bandgap close simultaneously, for every value of $\delta\theta$ (Supplementary Text S2 and S7). The additional boundaries at $\theta=\pm\pi/7$ mark a second, narrower reversal of $w$ that is not tied to any change in $n$; we find this feature traces back to the specific bar-length and hinge-offset ratios used in our numerical examples (Supplementary Text, Figs.~S4-S6) rather than to a universal geometric constant, and we caution against over-generalizing this particular threshold.

We regard the shared transition at $\theta=0$ as a structural consequence of this parameterization --- both invariants are ultimately governed by the same inversion-symmetry-breaking parameter --- rather than as an a priori coincidence; a general proof for arbitrary asymmetric-angle parameterizations remains open. The rotor chain thus provides an explicit example in which kinematic and dynamical topology remain distinct yet are tied to a common critical geometry. To our knowledge, this is the first mechanical realization of dual topological channels whose transitions are symmetry-linked while remaining independently programmable.

\subsection{Independent programming of the dynamical channel through axial-stiffness dimerization}
\begin{figure}[h]
	\centering
	\includegraphics[width=0.89\textwidth]{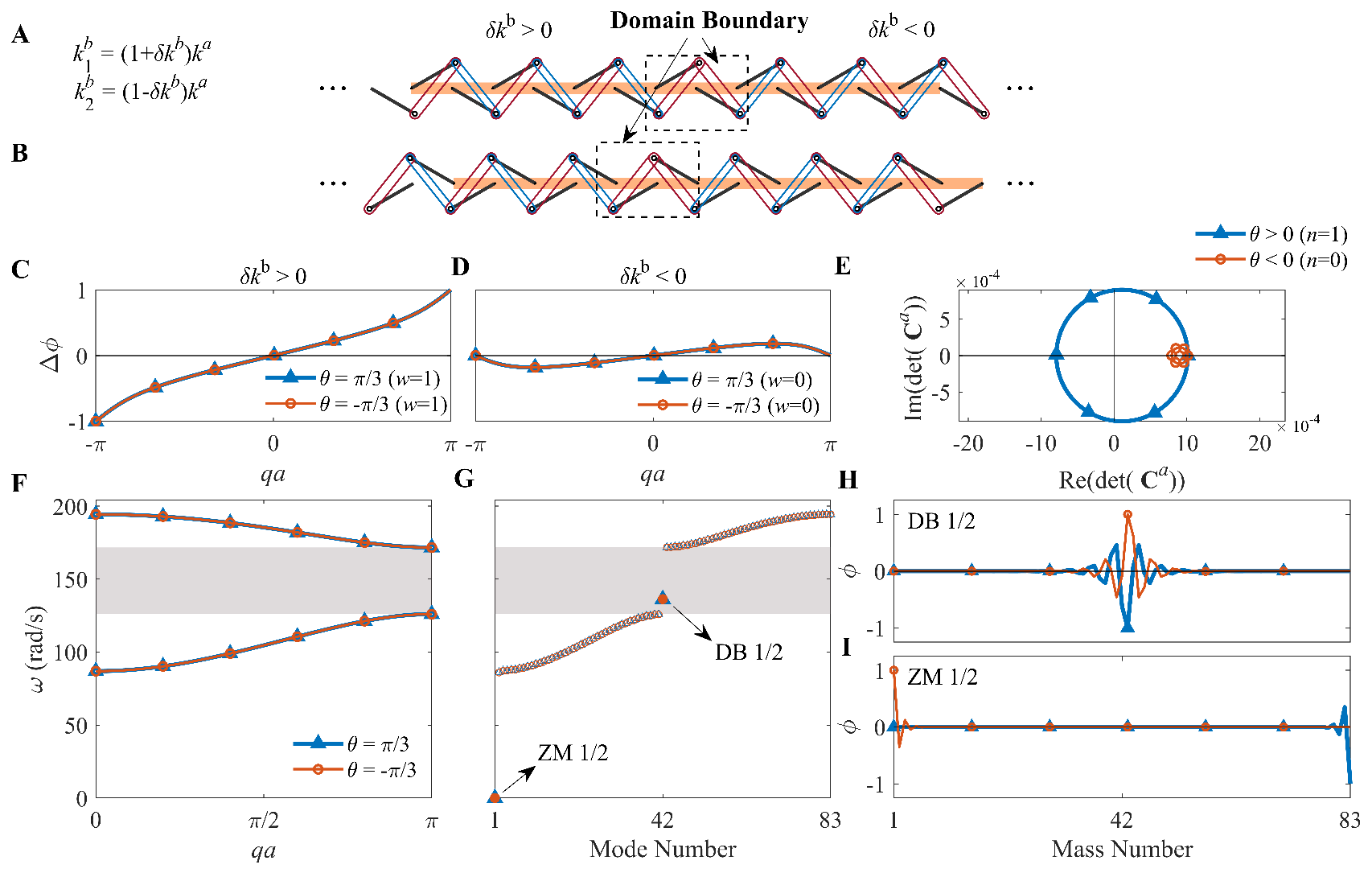}
	\caption{\textbf{Axial-stiffness dimerization switches the finite-frequency invariant independently of geometry.} (\textbf{A} and \textbf{B}) Finite lattices with DBs between domains of opposite $\delta k^b$ for $\theta>0$ (\textbf{A}) and $\theta<0$ (\textbf{B}). (\textbf{C} and \textbf{D}) Eigenvector phase difference confirming $w$ for $\delta k^b>0$ (\textbf{C}) and $\delta k^b<0$ (\textbf{D}). (\textbf{E}) Polarization winding trajectories of $\det(\mathbf{C}^a)$. (\textbf{F}) Dispersion showing the bandgap induced by $\delta k^b\neq0$. (\textbf{G}) Supercell spectrum showing DB and ZM states. (\textbf{H}) DB mode shapes for $\theta>0$ (blue) and $\theta<0$ (orange). (\textbf{I}) ZM mode shapes for (\textbf{A}) and (\textbf{B}), unchanged despite the new finite-frequency gap.}
	\label{Fig04}
\end{figure}

The shared critical geometry identified above does not imply that the two channels lose their independence. We next demonstrate that the dynamical channel can be switched without modifying the kinematic channel by introducing axial-stiffness dimerization while keeping the geometry fixed. Dimerizing the axial bond stiffness as $k_{1,2}^b = (1 \pm \delta k^b)k^a$ with $\delta k^b>0$ on one domain and $\delta k^b<0$ on the other, at fixed, symmetric pendulum angles ($\theta_{u,l}=\pm\pi/3$ and $\delta\theta=0$) (Fig.~\ref{Fig04}A and B), reproduces the full SSH phenomenology (Fig.~\ref{Fig04}): the finite-frequency invariant $w$ is determined strictly by the sign of $\delta k^b$, and entirely independent of the geometric parameter $\theta$ (Fig.~\ref{Fig04}C and D), while the zero-frequency polarization $n$ remains governed exclusively by $\theta$ and is unaffected by $\delta k^b$ (Fig.~\ref{Fig04}E) (Supplementary Text S4; Figs. S9–S11 give further examples).

Axial-stiffness dimerization therefore acts as a purely dynamical design parameter, providing an independent mechanism for switching the finite-frequency topological channel without altering the kinematic topology of the underlying frame.

\subsection{Rotational-stiffness dimerization lifts the zero mode without changing its topology}
\begin{figure}[h]
	\centering
	\includegraphics[width=0.89\textwidth]{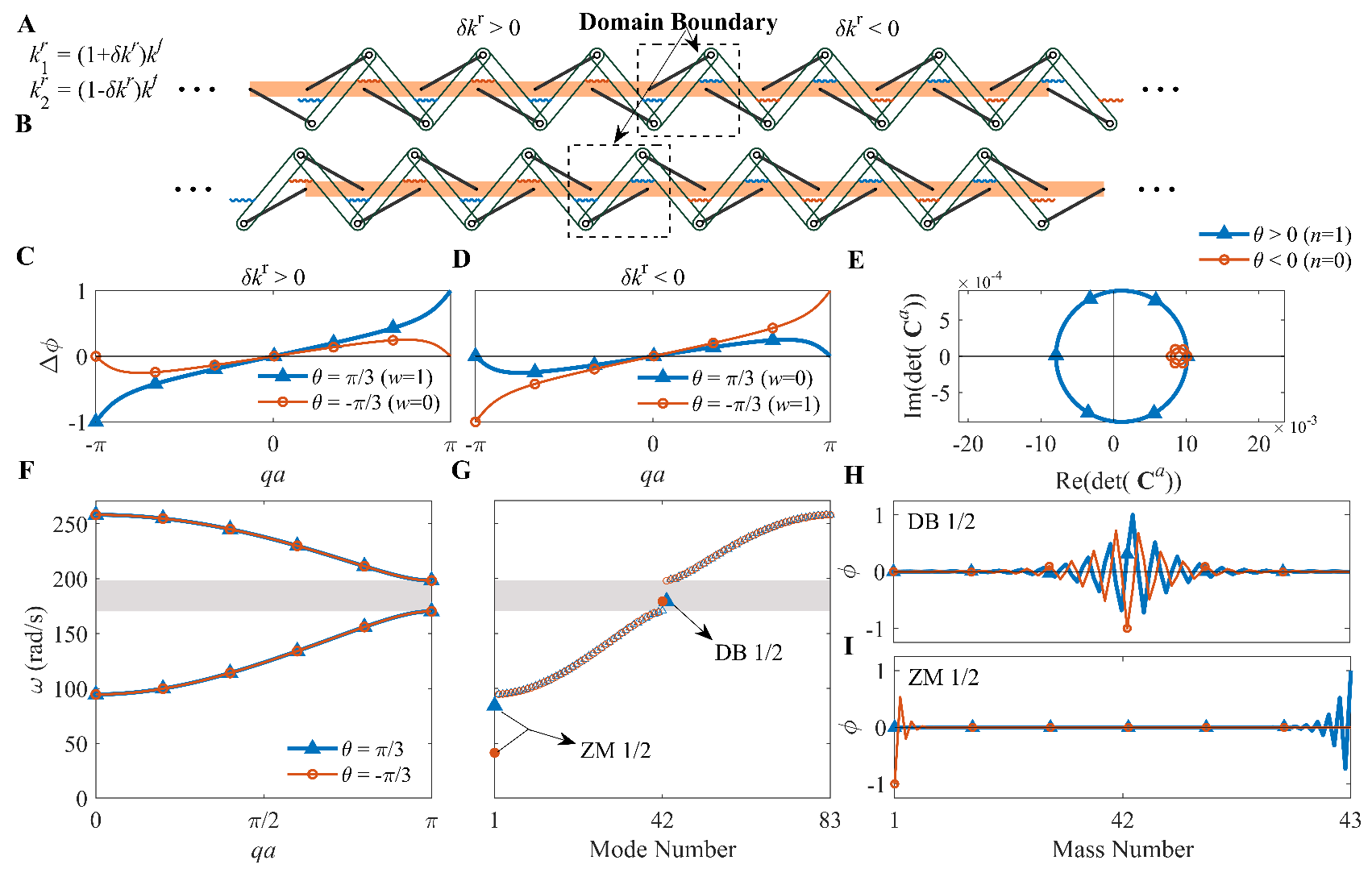}
	\caption{\textbf{Rotational-stiffness dimerization switches the finite-frequency invariant while lifting the ZM to a tunable finite frequency.} (\textbf{A} and \textbf{B}) Finite lattices with DBs between domains of opposite $\delta k^r$ for $\theta>0$ (\textbf{A}) and $\theta<0$ (\textbf{A}). (\textbf{C} and \textbf{D}) Eigenvector phase difference confirming $w$ for $\delta k^r>0$ (\textbf{C}) and $\delta k^r<0$ (\textbf{D}). (\textbf{E}) Polarization winding trajectories of $\det(\mathbf{C}^a)$. (\textbf{F}) Dispersion showing the bandgap induced by $\delta k^r\neq0$. (\textbf{G}) Supercell spectrum showing DB states and the lifted ZM. (\textbf{H}) DB mode shapes for $\theta>0$ (blue) and $\theta<0$ (orange). (\textbf{I}) Lifted ZM mode shapes for (\textbf{A}) and (\textbf{B}).}
	\label{Fig05}
\end{figure}

A complementary design pathway acts on the kinematic channel while simultaneously offering an additional, independent handle on the dynamical channel. In the limit of vanishing hinge stiffness, the floppy mode is protected by a continuous rotational symmetry and therefore resides at zero frequency. Introducing rotational stiffness --- physically, the bending resistance of a compliant crease or flexure ~\cite{Wu1,Ma3,sun1,zhang1} --- breaks this continuous symmetry and lifts the floppy mode to a finite, controllable frequency while preserving its topological character, which remains protected and governed exclusively by $\theta$ (Fig.~\ref{Fig05}).

Rotational stiffness therefore provides two independent design handles at once: introducing it lifts the kinematic mode to a tunable, protected finite frequency, while dimerizing it additionally switches a finite-frequency invariant within that lifted band --- both without altering the topological character set by $\theta$.

Dimerizing the rotational stiffness as $k_{1,2}^r = (1 \pm \delta k^r)k^f$ does more than shift the lifted mode's frequency: reversing the sign of $\delta k^r$ across an interface switches a finite-frequency invariant within the lifted band, generating a domain-boundary state in direct analogy with axial-stiffness dimerization (Supplementary Text S5; Figs. S12–S13). Physically, the lifting itself is a useful mechanical feature in its own right --- it stabilizes an otherwise marginal, zero-energy floppy mode against buckling or collapse under small perturbations --- and the magnitude of the frequency shift can be tuned by the competing axial and rotational energy scales (Supplementary Fig.~S5C).

Table~\ref{tab:design} summarizes how each of the four parameters discussed above --- the average angle $\theta_{avg}$, the angular asymmetry $\delta\theta$, and the two stiffness-dimerization ratios --- governs the two topological invariants, and which invariant each leaves unaffected.

\begin{table}[t]
	\centering
	\caption{Independent and joint control of the two topological channels.}
	\label{tab:design}
	\begin{tabular}{p{3.3cm}p{5.5cm}p{5.3cm}}
		\hline
		\textbf{Parameter} & \textbf{Primary effect} & \textbf{Left unchanged} \\
		\hline
		Average angle $\theta_{avg}=(\theta_u+\theta_l)/2$ &
		Sets the zero-frequency polarization $n$; also sets an exact finite-frequency transition at $\theta_{avg}=0$, independent of $\delta\theta$ &
		--- (governs both invariants at this shared point) \\[4pt]
		Angular asymmetry $\delta\theta$ &
		Sets the finite-frequency bandgap magnitude (Fig.~S6); jointly with $\theta$, sets the location of a second $w$-transition (near $\theta\approx\pm\pi/7$ for the parameters studied) &
		The $\theta_{avg}=0$ transition itself, and $n$ (unaffected by $\delta\theta$ at any $\theta$) \\[4pt]
		Axial stiffness ratio $k_1^b/k_2^b$ &
		Switches $w$ (sign of $k_1^b-k_2^b$), independent of $\theta$ &
		$n$ (remains set by $\theta_{avg}$) \\[4pt]
		Rotational stiffness ratio $k_1^r/k_2^r$ &
		Lifts the zero mode to a tunable finite frequency; dimerizing it also switches a finite-frequency invariant within the lifted band &
		Topological character of the lifted floppy mode (governed by $\theta_{avg}$) \\
		\hline
	\end{tabular}
\end{table}

\subsection{Experimental validation of dual topological channels}
\begin{figure*}[t]
		\centering
		\includegraphics[width=0.85\textwidth]{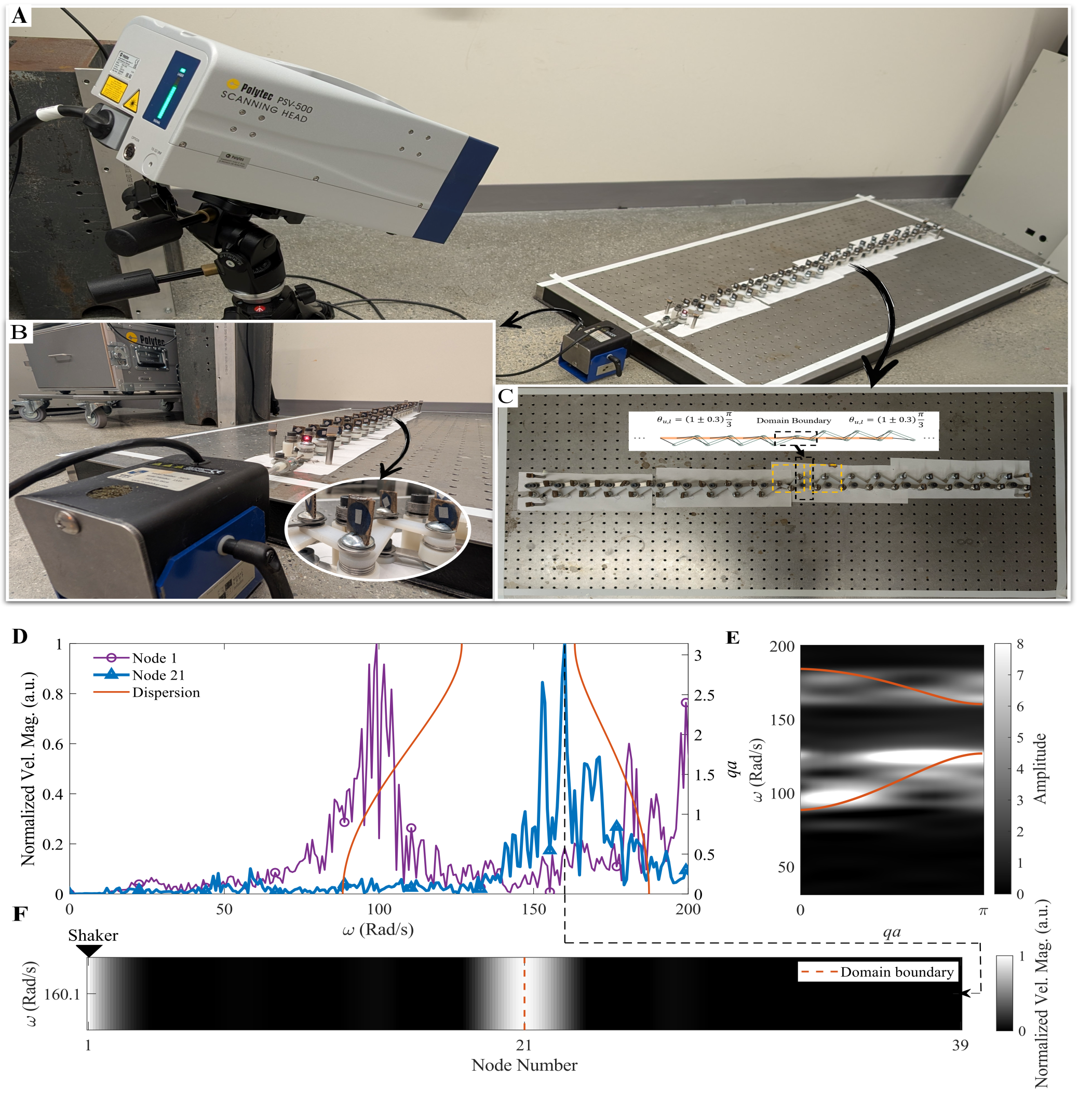}
		\caption{\textbf{Experimental validation of finite-frequency topological states.} (\textbf{A}) Experimental setup showing the scanning laser Doppler vibrometer (SLDV) measurement system. (\textbf{B}) Shaker excitation with a zoomed-in inset detailing the reflective markers used to project in-plane rotations onto the out-of-plane measurement axis. (\textbf{C}) Top view of the rotor-chain specimen, indicating the left and right domains and the central domain boundary. (\textbf{D}) Frequency response spectra measured at Node 1 (input) and Node 21 (domain boundary), illustrating the localized response within the finite-frequency bandgap. (\textbf{E}) 2D-DFT of the experimental spatio-temporal data overlaid with the theoretical band structure (orange solid lines), confirming the physical opening of the finite-frequency bandgap. (\textbf{F}) Spatial displacement profile across the lattice at the domain-boundary mode frequency ($\omega \approx 160.1$ rad/s), demonstrating strong wave localization at Node 21.}
		\label{Fig06}
	\end{figure*}

To test the theoretical predictions experimentally, we fabricated a 19-unit-cell rotor chain with a geometric domain boundary at base angle $\theta=\pi/3$ and angular asymmetry $\delta\theta=0.3$ (Fig.~\ref{Fig06}A; see Materials and Methods and Supplementary Text S10 for fabrication details). This configuration was chosen because it simultaneously supports a finite-frequency topological domain-boundary state and a boundary-localized floppy mode, thereby providing a direct test of the dual-channel framework developed above. Here we focus on experimentally validating both the finite-frequency channel induced by angular asymmetry and the zero-frequency floppy-mode localization predicted by the polarization invariant.

Under free--free boundary conditions, the chain was excited using a 20--50 Hz burst-chirp signal from an electrodynamic shaker. A scanning laser Doppler vibrometer recorded the out-of-plane velocity of reflective markers mounted perpendicular to each node, which project the in-plane pendulum rotations onto the measurement axis (Fig.~\ref{Fig06}A).

A two-dimensional discrete Fourier transform of the measured space--time velocity field maps the response into the frequency--wavenumber domain (Fig.~\ref{Fig06}B). The measured spectrum closely follows the analytically predicted acoustic and optical branches (root-mean-square error 8.0 rad/s on the acoustic branch and 9.3 rad/s on the optical branch, corresponding to 4.7\% and 5.5\% of the measured bandwidth, respectively) and clearly captures the opening of the finite-frequency bandgap induced by the angular asymmetry $\delta\theta$. These measurements confirm the existence of the predicted bulk wave-propagation channel.

The opening of a bulk bandgap alone, however, does not establish its topological character. To probe the associated boundary physics directly, we examine the spatial velocity profile at representative frequencies inside and outside the gap (Fig.~\ref{Fig06}C). Excitation within the propagating bands produces an extended response spanning the entire lattice, whereas excitation within the bandgap yields a sharply localized response centered at the domain boundary. This localization provides direct experimental evidence of the predicted finite-frequency topological domain-boundary state.

The complementary kinematic channel was assessed on the same specimen. Because an ideal floppy mode is associated with vanishing restoring stiffness, it can be identified through its highly compliant deformation pathway. Manual actuation of the boundary predicted by the polarization invariant ($n=1$) produced large-amplitude motion with negligible resistance, while the remainder of the lattice remained comparatively rigid. Repeating this test along the full chain confirmed that the compliant response is confined to the predicted boundary and neither propagates into the bulk nor appears at the opposite end (Supplementary Video S1; see Materials and Methods).

Together, the measured bulk dispersion, domain-boundary localization, and boundary-localized floppy mode confirm the central prediction of this work: a single mechanical architecture can simultaneously host a dynamical topological channel and a kinematic topological channel. The experiments therefore validate not only the coexistence of static and dynamic topological response, but also the broader dual-channel design strategy introduced in this study. 

	\section{Discussion}
Topological mechanical metamaterials have traditionally been divided into two largely separate classes: those whose robustness derives from kinematic compatibility and supports zero-frequency floppy modes, and those whose robustness derives from finite-frequency band topology and supports protected wave localization. Here we show that these two paradigms need not be mutually exclusive. By introducing a second pendulum degree of freedom into the classical rotor chain, we realize a single architecture that simultaneously hosts a kinematic topological channel and a dynamical topological channel. Moreover, these channels remain independently programmable despite coexisting within the same unit cell, enabling static and dynamic response to be controlled separately.

A central result of this work is that the two channels are not merely co-located, but are connected through a shared geometric control parameter. The kinematic channel is governed by the average pendulum orientation and determines the localization of the zero-frequency floppy mode. The dynamical channel, by contrast, emerges from finite-frequency band topology enabled by inversion-symmetry breaking within the unit cell. Despite these distinct physical origins, the corresponding topological transitions coincide at the same critical geometry under the parameterization considered here. Specifically, both transitions occur when the average angle passes through $\theta_{avg}=0$, where the unit cell recovers inversion symmetry and both the polarization gap and the finite-frequency bandgap close simultaneously. The rotor chain therefore provides an explicit example in which geometric and band topology remain distinct yet become locked to a common critical geometry.

Our findings complement recent efforts to extend Maxwell topology beyond the zero-frequency limit. Previous studies have shown that introducing elastic stiffness can lift topological floppy modes to finite frequency while preserving their underlying topological character~\cite{Ma3,sun1,zhang1}, and related finite-frequency Maxwell modes have been demonstrated in self-dual kagome lattices~\cite{danawe1}. In these systems, however, a single topological invariant persists across different frequency regimes. By contrast, the present rotor chain supports two distinct invariants simultaneously. The geometric polarization and finite-frequency band topology remain separately identifiable and can be manipulated through different design parameters without disrupting one another. This separation introduces a level of functional programmability that is absent from existing topological mechanical architectures.

The coexistence of independently programmable static and dynamic channels suggests new opportunities for multifunctional metamaterials. In vibration-mitigation systems, for example, a protected finite-frequency stopband could be reconfigured without altering the structure's static compliance or load-bearing pathway. More broadly, the combination of a floppy-mode channel and a wave-guiding channel enables architectures in which quasi-static deformation and dynamic signal transmission are controlled independently within the same physical platform. Such capabilities could support mechanically reconfigurable waveguides, adaptive vibration isolators, mechanical logic elements, and sensing systems that exploit topological robustness across multiple operating regimes.

Although demonstrated here in a one-dimensional rotor chain, the underlying concept is substantially more general. The dual-channel framework introduced in this work does not rely on one-dimensionality and should extend naturally to higher-dimensional Maxwell lattices~\cite{Ma3,zhang1}, hierarchical~\cite{Pyfrom1} and long-range-coupled~\cite{Rajabpoor2} phononic systems, and non-centrosymmetric photonic, acoustic, and electronic analogs~\cite{Silva1, Harris1, Dorgant1, Shen1}. The incorporation of nonlinear, active, or programmable elements~\cite{Zhou1, Li2, Trainiti1, Xiu1, Chen1, Wang3} may further enable the two channels to be switched independently and reversibly in situ. More broadly, our results demonstrate that kinematic and dynamical topology need not be treated as separate design paradigms but can instead be combined within a unified framework for engineering independently programmable static and dynamic response. By establishing dual topological channels as a design principle rather than a property of a specific lattice, this work opens a pathway toward multifunctional topological systems whose mechanical behavior can be programmed across both static and dynamic regimes.

\section*{Materials and Methods}
\subsection*{Analysis and numerical framework}
The Bloch compatibility and stiffness matrices, the definitions of the zero- and finite-frequency winding numbers, and their closed-form dependence on geometric and stiffness parameters are derived in full in Supplementary Text S1 to S5. Bulk dispersion curves were obtained by diagonalizing the Bloch stiffness matrix over the first Brillouin zone; domain-boundary and zero-frequency states were obtained by diagonalizing the stiffness matrix of a finite supercell (Supplementary Text S1) under free boundary conditions.

\subsection*{Fabrication}
The bulk lattice and interface region were cut from ultra-high-molecular-weight polyethylene (UHMWPE, $E = 750$ MPa) sheet, with bar widths individually adjusted across four element types to hold an approximately uniform effective axial stiffness despite their different lengths (Supplementary Text S9, Table S1). Joints used steel binding barrel screws ($E = 710$ GPa) with localized lubrication to approximate frictionless hinges; shoulder screws, washers, and plastic spacers maintained planar alignment. Hardware and bar mass were partitioned to each node, giving an effective nodal mass of 40 g used in the analytical mass matrix (Supplementary Text S9).

\subsection*{Vibrometry measurement}
The 19-unit-cell specimen ($\theta = \pi/3$, $\delta\theta = 0.3$) was suspended under free--free boundary conditions and excited with a 20--50 Hz burst-chirp signal from an electrodynamic shaker, with the burst spanning 70\% of a 4-s acquisition window to allow steady-state response while leaving sufficient free-decay time to avoid spectral leakage. Reflective markers mounted perpendicular to each node projected in-plane pendulum rotation onto the out-of-plane axis measured by a scanning laser Doppler vibrometer. A two-dimensional discrete Fourier transform of the resulting space--time velocity field yielded the frequency--wavenumber spectrum shown in Fig.~\ref{Fig06}e, and the spatial velocity profile at the domain-boundary mode frequency ($\omega = 160.1$ rad/s) yielded the localization comparison shown in Fig.~\ref{Fig06}f. The spatial velocity magnitude was normalized against the maximum measured amplitude, confirming strong energy confinement at the domain boundary without the need for an additional quantitative localization metric.

\subsection*{Quasi-static floppy-mode test}
The zero-frequency channel was probed on the same specimen by manually displacing the pendulums at each end of the chain and recording the response with a handheld camera at normal frame rate (no high-speed or slow-motion capture, consistent with the quasi-static nature of the test). At the boundary predicted to host the floppy mode, the pendulums rotated with negligible resistance; the remaining unit cells showed no perceptible motion. This test was repeated moving cell by cell along the chain to confirm that the low-resistance response is confined to the predicted boundary and is absent elsewhere, including the opposite end. This is a qualitative demonstration of mode localization rather than a quantitative stiffness or force measurement (Supplementary Video S1).

\section*{Acknowledgments}

\noindent\textbf{Funding:} 

\noindent\textbf{Author contributions:} 

\noindent\textbf{Competing interests:} 

\noindent\textbf{Data, code and materials availability:} 
	\bibliography{References_SA}
\bibliographystyle{sciencemag}

\clearpage
\includepdf[pages=-]{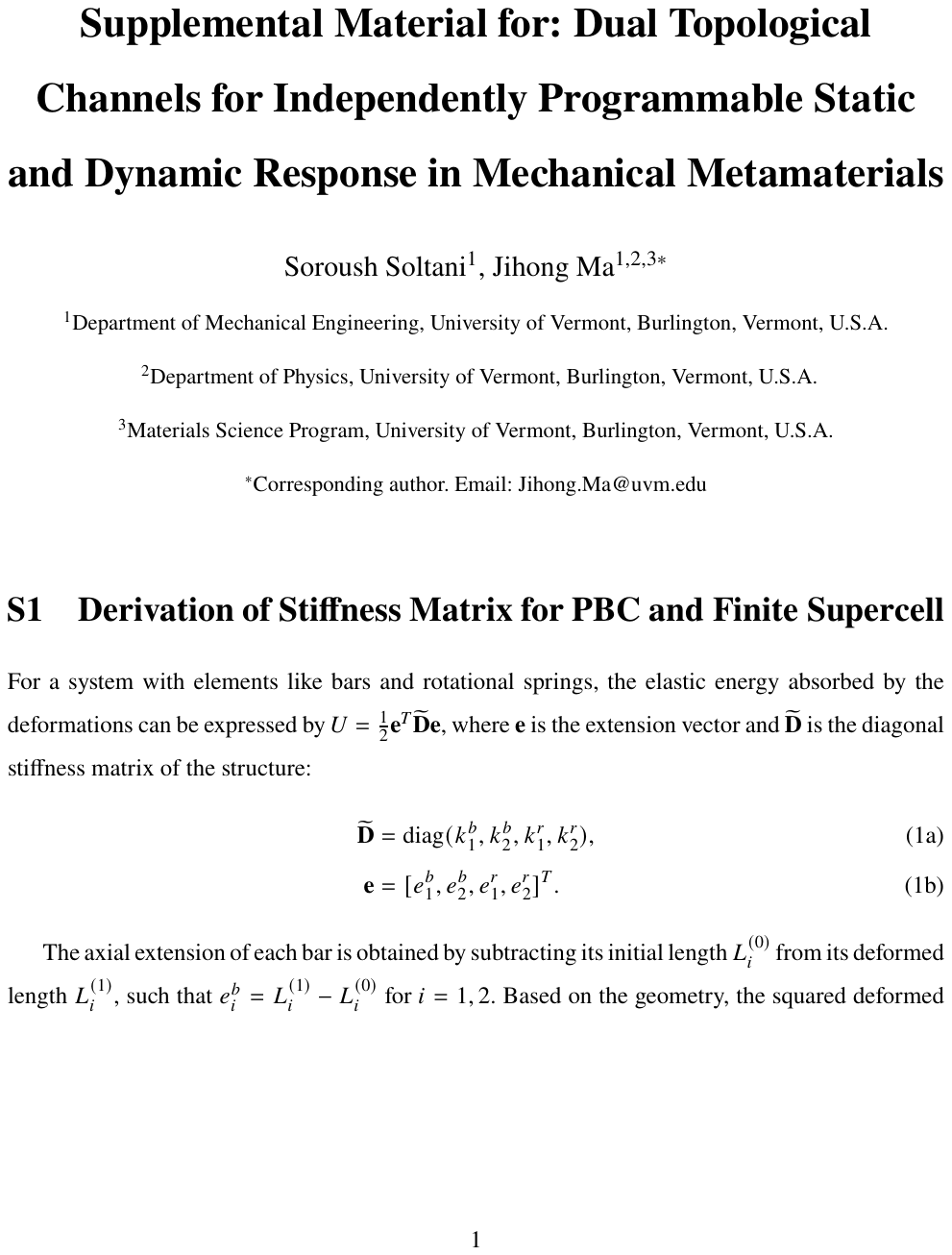}

\clearpage
\includepdf[pages=-]{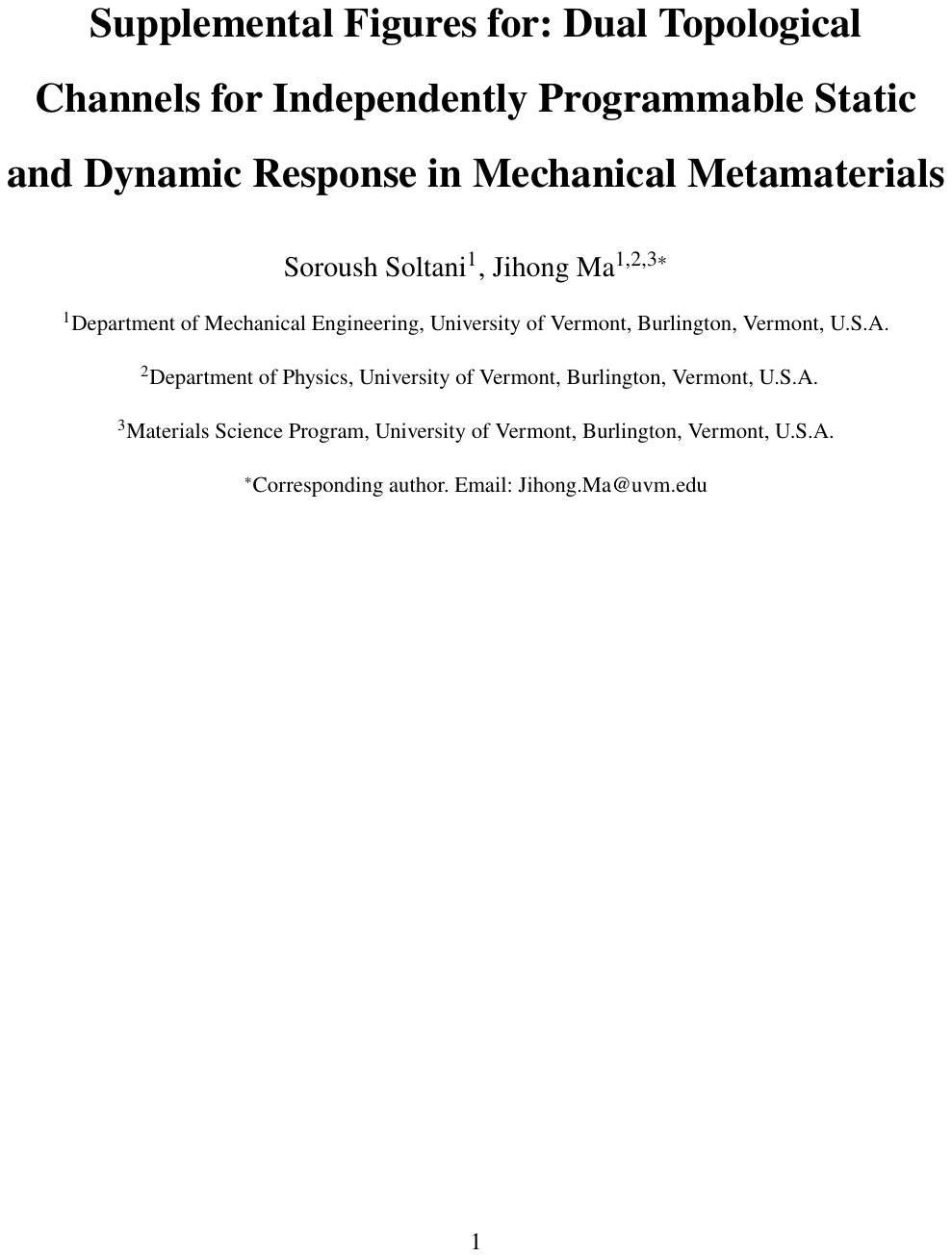}

\end{document}